# Vector rectangular-shape laser based on reduced graphene oxide interacting with long fiber taper

Lei Gao, Tao Zhu, *Member, IEEE*, Jing Zeng, Wei Huang, and Min Liu

*Abstract*—A vector dual-wavelength rectangular-shape laser (RSL) based on a long fiber taper deposited with reduced graphene oxide is proposed, where the nonlinearity is enhanced due to large evanescent-field-interacting length and strong field confinement of a 8 mm fiber taper with a waist diameter of 4 μm. Graphene flakes are deposited uniformly on the taper waist with light pressure effect, so this structure guarantees both excellent saturable absorption and high nonlinearity. The RSL with a repetition rate of 7.9 MHz exhibits fast polarization switching in two orthogonal polarization directions, and the temporal and spectral characteristics are investigated. The results suggest that the long taper-based graphene structure is an efficient choice for nonlinear devices.

## I. Introduction

GRAPHENE has received considerable attentions as mode-locker for passively mode-locked fiber lasers (MLs), which have wide applications in both fundamental and industrial applications, such as nonlinear optics, micro-machining, communication and sensing [1-3]. As a two-dimensional layer of carbon atoms with hexagonal packed structure, the Pauli blocking caused by the linear dispersion of the Dirac electrons provides it excellent saturation absorption virtues, including broad bandwidth, large modulation depth, and low saturation intensity [3-5]. Recently, several groups found that the 3rd-order susceptibility of graphene is ultrahigh [6-8], so it can be utilized in nonlinear devices and systems. Luo *et al* observe the graphene-induced nonlineaer four-wave-mixing, which can be used for multiwavelength fiber laser [9]. However, the high nonlinearity has not been fully explored. Here, the nonlinearity of graphene is utilized for generating rectangular-shape laser (RSL) for the first time.

Behaving as square shape trains temporally, RSL is a potential optical source in telecommunication and fiber Bragg grating interrogation [10-12]. Most of the reported RSLs based on the nonlinear polarization rotation (NPR) or semiconductor saturable absorber mirrors (SESAMs) have characteristics of broaden temporal duration with fixed amplitude when increasing pump power, which is explained by dissipative soliton resonance (DSR) [13-15]. Other schemes generating RSL would require a very long length of high nonlinear fibers [10,16].

In this letter, we illustrate that the high nonlinearity of reduced graphene oxide can be utilized for RSL with easiness. The saturable absorber (SA) is produced by depositing graphene flakes on a long fiber taper that inserted in an erbium-doped fiber (EDF) cavity with anomalous group velocity dispersion (GVD). Although the ~8 mm taper waist length is much shorter than that of the reported nonlinear fibers (normally several hundreds of meters) [10,16], it can enhance the cavity nonlinearity due to large evanescent-field-interacting length, strong field confinement and high nonlinearity of graphene, which is key for generating RSL.

## II. System Configuration and Signal Processing

The reduced graphene oxide (rGO) used in our experiment is synthesized by reducing graphene oxide, and the thickness is 0.55~3.74 nm. Two major peaks in the Raman spectroscopy shown in Fig. 1 (a) corresponding to ~1320 cm$^{-1}$ and ~1610 cm$^{-1}$, respectively. To fabricate SA with low scattering loss, we immerse 0.5 mg rGO into 50 mL N,N-dimethylformamide solution for 30 minutes. The solution is centrifuged to separate transparent part, as shown in Fig. 1(a). A standard single mode fiber (SMF, Corning SMF-28) is tapered with low loss and good repeatability. The waist diameter and the length of the taper is ~4 μm and 8 mm, respectively. Compared with the reported taper length [17-19], the 8 mm taper waist in our laser cavity provides enough interacting length between the rGO and the transmission light, guaranteeing excellent saturable absorption and high nonlinearity. As shown in Fig. 1 (b), the insertion loss is ~ 2.7 dB, and slight interference can be observed due to the tapering process. After immersing the taper into a droplet of rGO solution, a 2 mW continuous 1550 nm laser is injected into one port of the taper and the output power from the other port is monitored. Due to light pressure effect, the rGO can deposit on the surface of the taper. One minute later, the output power drops ~2 dB, then the taper is removed from rGO solution and fixed in a clean box to dry.

This work was supported by National Natural Science Foundation of China (No. 61377066), and the Fundamental Research Funds for the Central Universities (No. CDJZR12125502, 106112013CDJZR120002 and 106112013CDJZR160006). Assistances of Tongtao Li and Zhe Luo from Chongqing University are appreciated.

Lei Gao, Tao Zhu, Jing Zeng, Wei Huang and Min Liu are with the Key Laboratory of Optoelectronic Technology & Systems (Education Ministry of China), Chongqing University, Chongqing 400044, China. (Corresponding email: zhutao@cqu.edu.cn)

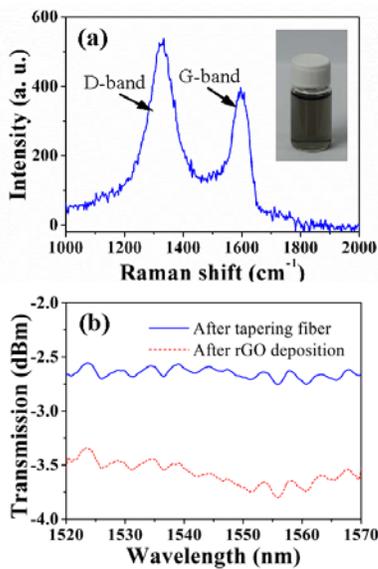

Fig. 1 (a), Raman spectrum of rGO sample excited with a 632.8 nm laser, and the inset shows the transparent solution of rGO and N,N-dimethylformamide. (b), transmission spectra of the fiber taper before and after rGO deposition.

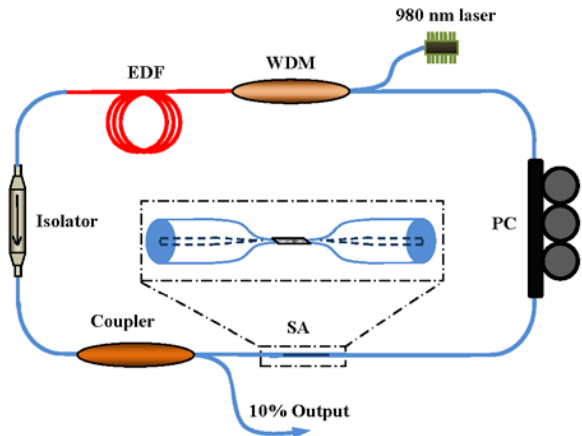

Fig. 2 Schematic diagram of the fiber ring laser.

Figure 2 shows the setup of fiber cavity incorporating the SA fabricated above. The EDF (EDF-980-T2, Stockeryale, Inc.) is 9.7 m with a peak absorption of 6.56 dB/m at 1530 nm and the laser is pumped by a 980 nm laser through a wavelength division multiplexer (WDM). A polarization independent isolator is used for unidirectional operation, and a polarization controller (PC) is employed to optimize the birefringence of the cavity. The output laser is extracted from a 10% fiber coupler. All other fibers used in the cavity are standard SMF. The GVD parameters of EDF and SMF are -12.7 and 18 ps/nm/km, respectively, corresponding to a net anomalous dispersion of 0.0155 ps2. The total length of the ring cavity is about 26.7 m, corresponding to a fundamental frequency of ~7.9 MHz. The laser output is monitored by a detector (PDB430C, 350MHz, Thorlabs Co,. Ltd), a real time oscilloscope (Infiniium MSO 9404A, 4GHz, Agilent Tech.), a frequency analyzer (DSA815, Rigol Tech., INC.) and an optical spectrum analyzer (Q8384, Advantest Corp.). Besides, a variable optical attenuator is inserted to make sure the detector is unsaturated.

III. EXPERIMENTAL RESULTS AND DISCUSSION

Increasing pump power with a step of ~5 mW, the output power is recorded and shown in Fig. 3 (a), where self-pulsing and RSL occur when pump power exceeds 18 mW and 179 mW, respectively. Figure 3 (b) shows a typical oscilloscope trace of the RSL with a fundamental period of 126.1 ns, and no fine structure can be observed by using a high bandwidth photo detector. Its down part has nonzero value, which is a result of the coupling of two wavelengths, and the inset reveals that the RSL has uniform shapes. The spectrum shown in Fig. 3 (c) contains two wavelengths centered at 1564.6 nm and 1567.4 nm, and the corresponding radio frequency (RF) spectrum in Fig. 3 (d) displays the wide RF spectrum up to 350 MHz. the fundamental repetition rate of ~7.9 MHz and ~60 dB of peak-to-background ratio are shown in the inset. Due to the Fourier transformation, two small sidebands are shown in each of the harmonic frequencies, but no pedestal can be found using a resolution of 100 Hz. Hence, we can see that this RSL exhibits good mode locking stability.

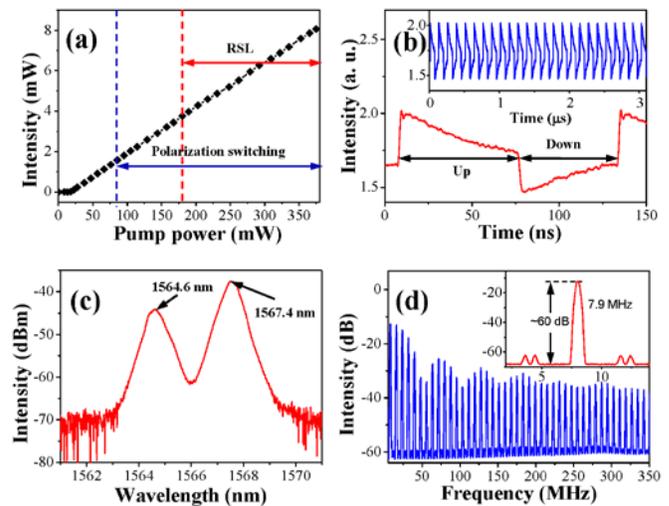

Fig. 3(a), Output intensities under different pump powers. (b), temporal trains for pump power at 380 mW, and the inset shows signals in a larger region. (c), optical spectrum corresponding to (b). (d), RF spectrum of laser at pump power of 380mW ( the inset shows the fundamental frequency is ~ 7.9 MHz).

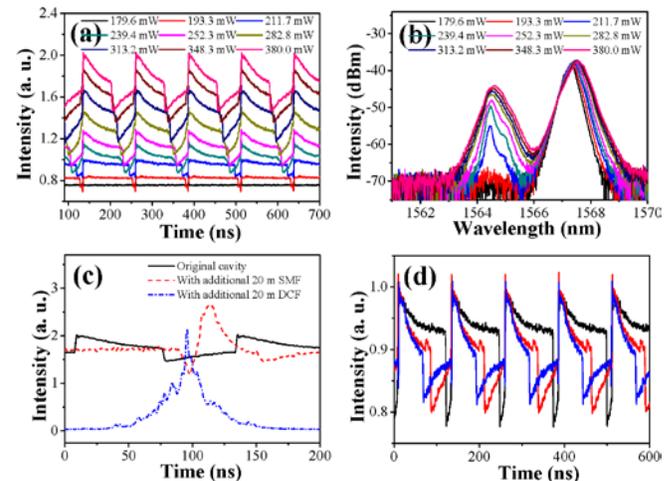

Fig. 4 (a), Temporal trains and (b), corresponding spectra for different pump powers. (c), Temporal trains in cavities with different GVDs. (d), Temporal trains under different PC biases.

The pump-dependent evolutions of the RSL are shown in Figs. 4 (a)-(b). For pump below 179 mW, no RSL is shown and only a single wavelength is presented. Continuously increasing pump strength results in RSL and dual wavelengths simultaneously. Contrary to square pulses caused by DSR, our rectangular pulse width narrows gradually when increasing the pumping strength, and its peak power increases approximately linearly. Meanwhile, its two wavelengths are broadened and red-shifted, which illustrates that this RSL is not DSR.

By adding different lengths of SMF and dispersion compensation fiber (DCF38, Thorlabs Co., Ltd) into our cavity to get different GVDs, we find that RSL exists only in cavity with anomalous GVD. Results in Fig. 4 (c) reveals that net normal GVD would produce conventional laser pulse, and larger net anomalous GVD would smooth the RSL shape. The temporal durations can be tuned by rotating the orientation of PC for diverse linear cavity phase delay bias settings, which can be seen in Fig. 4 (d).

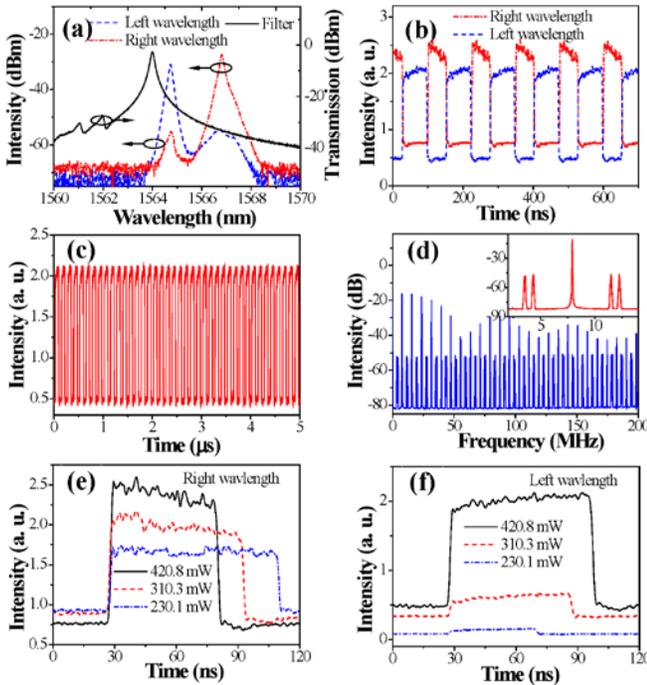

Fig. 5. (a), Spectra of two wavelengths separated by a tunable filter. (b), oscilloscope traces of the two separated wavelengths after the filter. (c) is the temporal trains of the left wavelength and (d) is the corresponding RF spectrum. (e) and (f) are the temporal components of the two wavelengths under different pump intensities after separated by the filter.

After the output laser is resolved by a tunable filter (FFP-TF2, Micron Optics), we found that the two wavelengths corresponding to two rectangular pulses with different intensities and durations, respectively. Limited by bandwidth of the filter, the separation of the two wavelengths is incomplete, yet it still can be found from Figs. 5 (a) and (b) that there are two pulses coexisting in the laser cavity, and the pulses with central wavelengths at ~1564.6 nm and ~1567.6 nm are responsible for the up and bottom durations, respectively. According to Figs. 5 (c) and (d), the RSL has been mode-locked well in each of the two wavelength. Figures 5 (e) and (f) show the two separated temporal trains with different pump intensities, where their durations varies inversely. It can be proved that the RSL is the result of the coupling of two lasing wavelengths that corresponds to each rectangular pulse.

The vector nature of the RSL is examined with a polarization beam splitter (PBS), and the two output ports are recorded by the two identical photo-detectors simultaneously. As shown in Fig. 6, the two wavelengths share almost the same polarization state, whereas their temporal outputs exhibit fast polarization switching dynamics [20]. This anti-phase polarization dynamics have been reported in continuous wave or MLs, which may result from the cross saturation and gain sharing of the two polarization axis, however, a clear physical interpretation is still controversial [20-23]. In our experiment, the polarization dynamics effect can be found only in cavity with net anomalous GVD, and it appears in both self-pulsing and RSL region. The detail characteristics of this effect will be our future work.

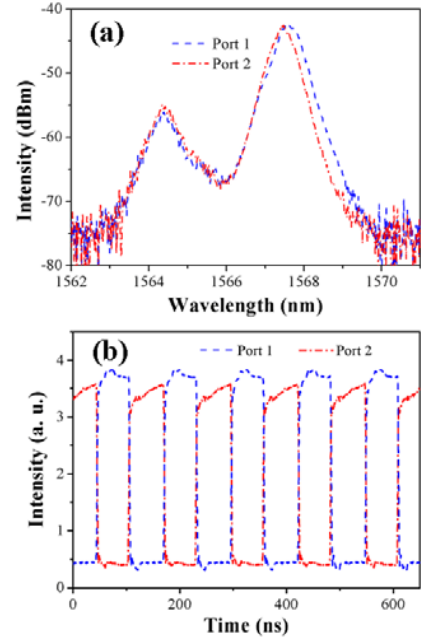

Fig. 6. (a), Spectra and (b) temporal trains from two ports of a PBS for pump power at 380 mW.

Compared with RSL generated by modulating a continuous wave laser via an optical modulator, our schemes mainly posses two main advantages. The first one is that the RSL in our experiment is a vector laser. As shown in Fig. 6 (b), the temporal trains of the two polarization directions almost show a complete switching in less than 1 ns. This kind of vector laser can be utilized as optical source for polarization division multiplexing for telecommunication, nanoparticles manipulation et al [24-26]. The second one is that two wavelengths are formed in our experiment, and the wavelength number is determined by the continuous wave laser source when using a conventional modulator. This kind of RSL with two wavelengths may find potential applications in specific circumstances.

## IV. CONCLUSION

We propose a vector dual-wavelength RSL based on long fiber taper deposited by rGO. The temporal and spectral characteristics with various pump powers are investigated, and results show that the RSL is due to the coupling between the two wavelengths. The long interacting length between the rGO and the transmission light guarantees excellent saturable absorption and high nonlinearity. This kind of RSL would find potential applications in telecommunication, sensing and grating interrogation. Our experiments suggest that the long taper-based graphene structure is an efficient choice for nonlinear devices or nonlinear systems.